\documentclass[conference,letterpaper]{IEEEtran}
\usepackage[utf8]{inputenc} 
\usepackage[T1]{fontenc}
\usepackage{url}
\usepackage{ifthen}
\usepackage{cite}
\usepackage[cmex10]{amsmath}

\usepackage{bbm}
\usepackage{amssymb}
\usepackage[shortlabels]{enumitem}
\usepackage{xcolor}
\usepackage{thmtools}
\usepackage{graphicx}
\usepackage[capitalise]{cleveref}

\declaretheorem[thmbox=M]{theorem}
\declaretheorem[thmbox=M]{proposition}
\declaretheorem[thmbox=M]{definition}
\declaretheorem[thmbox=M]{lemma}
\declaretheorem[thmbox=M]{corollary}
\declaretheorem[thmbox=M]{example}
\declaretheorem[style=remark]{remark}

\def \X{\mathsf X}
\def \Y{\mathsf Y}
\def \F{\mathcal F}
\def \calP{\mathcal P}
\def \calS{\mathcal S}
\def \calH{\mathcal H}
\def \reals {\mathbb R}
\def \supp {\mathrm{supp}}
\def \dualK{K^\star_\mu}
\newcommand{\renyiDiv}[2]{D_{\alpha}\left(#1\|#2\right)}
\newcommand{\dd}[1]{\mathrm{d}#1}
\newcommand{\mean}[2]{\mathbb{E}_{#1}\left[#2\right]}
\newcommand{\ip}[2]{\left\langle#1, #2\right\rangle}
\newcommand{\hellingerInt}[2]{H_{\alpha}\left(#1\|#2\right)}
\begin{document}
\title{Contraction of R\'enyi Divergences for Discrete Channels: Properties and Applications} 

\author{%
  \IEEEauthorblockN{Adrien Vandenbroucque}
  \IEEEauthorblockA{EPFL\\ 
                    Lausanne, Switzerland\\
                    adrien.vandenbroucque@epfl.ch}
  \and
  \IEEEauthorblockN{Amedeo Roberto Esposito}
  \IEEEauthorblockA{Okinawa Institute of Science and Technology\\
                    Onna, Japan\\
                    amedeo.esposito@oist.jp}
    \and
    \IEEEauthorblockN{Michael Gastpar}
    \IEEEauthorblockA{EPFL\\ 
                    Lausanne, Switzerland\\
                    michael.gastpar@epfl.ch}
}

\maketitle

\begin{abstract}
   This work explores properties of Strong Data-Processing constants for R\'enyi Divergences. Parallels are made with the well-studied $\varphi$-Divergences, and it is shown that the order $\alpha$ of R\'enyi Divergences dictates whether certain properties of the contraction of $\varphi$-Divergences are mirrored or not. In particular, we demonstrate that when $\alpha>1$, the contraction properties can deviate quite strikingly from those of $\varphi$-Divergences. We also uncover specific characteristics of contraction for the \mbox{$\infty$-R\'enyi Divergence} and relate it to \mbox{$\varepsilon$-Local Differential Privacy}. The results are then applied to bound the speed of convergence of Markov chains, where we argue that the contraction of R\'enyi Divergences offers a new perspective on the contraction of \mbox{$L^\alpha$-norms} commonly studied in the literature.
   \end{abstract}
\begin{IEEEkeywords}
    R\'enyi Divergences, $\varphi$-Divergences, Data-Processing Inequality, Strong Data-Processing Inequality, Markov Chains, Local Differential Privacy
\end{IEEEkeywords}

\section{Introduction}\noindent
The data-processing inequality (DPI) is a fundamental result in information theory, formalising the fact that post-processing cannot increase ``information''. In the last decade, efforts have been made to better understand a tightening of DPI known as the Strong Data-Processing Inequality (SDPI), whose purpose is to quantify the information contraction guaranteed by the DPI~\cite[Chapter 33]{Polyanskiy2025}. A growing body of work addresses both the computation of SDPI constants, their properties, and the inequalities governing their relationships. Part of the attractiveness comes from the applicability of such inequalities; they find practical use in information-theoretic settings such as local differential privacy~\cite{balle2019privacy,asoodeh2020contraction}, information percolation, broadcasting on trees, and information reconstruction~\cite{polyanskiy2017strong,polyanskiy2020application,gu2023non}, but also in the study of Markov processes or concentration inequalities, and their recent applications in machine learning~\cite{esposito2024concentration,koloskova2025certified,chien2024langevin}.

SDPIs for $\varphi$-Divergences have been thoroughly studied, and it is now established that a variety of contraction properties are common to all of them~\cite{raginsky2016strong}. In this work, we restrict our attention to the contraction of R\'enyi Divergences of order $\alpha\geq 0$, a setting for which results are currently sparse~\cite{jin2024properties,abawonse2025generalized,grosse2025bounds}. We show that depending on the order of the R\'enyi Divergences, certain contraction properties characteristic of \mbox{$\varphi$-Divergences} may or may not be recovered. Differences can manifest in drastic ways; for instance, channels may be viewed as contractive under \mbox{$\varphi$-Divergences} but non-contractive under R\'enyi Divergences. After briefly connecting the contraction of the $\infty$-R\'enyi Divergence to Local Differential Privacy (LDP), we apply R\'enyi's SDPIs to bound how fast Markov chains converge to their stationary distribution. In that context, a fresh perspective on bounds provided by R\'enyi Divergences is highlighted, namely that they can be seen as non-linear contractions of $L^\alpha$-norms.

The remainder of the paper is organised as follows. \Cref{section:background_and_definitions} presents the necessary background, including notation and definitions of divergences and SDPIs. We then proceed by establishing various bounds and identities pertaining to SDPIs for R\'enyi Divergences in~\cref{section:main_results}. Finally, a connection to $\varepsilon$-LDP is  made in~\Cref{section:applications}, followed by applying our result in the study of convergence of Markov chains.

\section{Background and Definitions}\label{section:background_and_definitions}\noindent
In the following, we consider spaces with finite alphabets. For such a space $\X$, it is assumed that a $\sigma$-algebra $\Sigma_\X$ is associated to it, rendering it a measure space. Since the considered spaces are finite, we can assume $\Sigma_\X$ to be the power set of $\X$. Indeed, in case it is not, some symbols in $\X$ can be merged to yield a new measure space with this property. We denote by $\calP(\X)$ the set of probability measures on $\X$ and by $\F(\X)$ the set of all real-valued functions on $\X$. The expectation of a function $f\in\F(\X)$ w.r.t. a probability measure $\mu\in\calP(\X)$ is ${\mean{\mu}{f}\triangleq\sum_{x\in\X}f(x)\mu(x)}$ and its $L^\alpha$-norm is ${\|f\|_{L^\alpha(\mu)}\triangleq\big(\mean{\mu}{|f|^\alpha}\big)^{\frac1\alpha}}$ for $\alpha\geq1$. Given two probability measures $\nu, \mu$, we write $\nu\ll\mu$ to denote that $\nu$ is absolutely continuous w.r.t. $\mu$, in which case there exists a Radon-Nikodym derivative $\frac{\dd\nu}{\dd\mu}$. The natural logarithm function is denoted by $\log$. The Binary Symmetric Channel with crossover probability $\varepsilon$ is written as $\mathrm{BSC}(\varepsilon)$.

\subsection{Markov Kernels}\label{sec:markov_kernels}\noindent
Hereafter, a channel will be viewed as a Markov kernel, or stochastic transformation.
\begin{definition}
    A Markov kernel $K:\Y\times\X\to[0, 1]$ specifies transition probabilities from elements of a set $\X$ to elements of a set $\Y$, and in particular
\begin{enumerate}
    \item $\forall x\in\X$, the mapping $y\mapsto K(y|x)$ is a probability distribution,
    \item $\forall y\in\Y$, the mapping $x\mapsto K(y|x)$ is a $\Sigma_\X$-measurable real-valued function.
\end{enumerate}
\end{definition}

A Markov kernel acts on probability distributions $\mu\in\calP(\X)$ as
\begin{equation*}
    \mu K (y) = \sum_{x\in\X} K(y|x)\mu(x), \quad \text{for }y\in\Y,
\end{equation*}
so that $\mu K\in\calP(\Y)$.
The set of all such kernels is denoted by $\calP(\Y|\X)$. 

\subsection{Divergences and (Strong) Data-Processing Inequalities}\noindent
Let us define two families of divergences which will be central to this work, starting with R\'enyi Divergences~\cite{van2014renyi}.
\begin{definition}
    For two probability measures $\nu, \mu \in\calP(\X)$ such that $\nu \ll \mu$, the R\'enyi divergence of order $\alpha \geq 0$ from $\nu$ to $\mu$ is defined as 
    \begin{equation*}
        \renyiDiv{\nu}{\mu} = \frac{1}{\alpha - 1}\log\mean{\mu}{\left(\frac{\dd\nu}{\dd\mu}\right)^{\alpha}},
    \end{equation*}
    where the orders $0, 1$ and $\infty$ are defined by continuous extensions.
\end{definition}
The KL Divergence, denoted as $D_{\rm KL}(\nu\|\mu)$, corresponds to the limit $\alpha\to1$.

The second family of divergences we consider are the $\varphi$-Divergences~\cite{csiszar1967information}.
\begin{definition}
    Let $\varphi:\reals_+\to\reals$ be a convex function with $\varphi(1)=0$. For two probability measures $\nu, \mu \in\calP(\X)$ such that $\nu \ll \mu$, the $\varphi$-Divergence from $\nu$ to $\mu$ is defined as
    \begin{equation*}
        D_\varphi(\nu\|\mu) = \mean{\mu}{\varphi\left(\frac{\dd\nu}{\dd\mu}\right)}.
    \end{equation*}
\end{definition}
A variety of well-known divergences belong to the family of $\varphi$-Divergences, notably:
\begin{itemize}
    \item The KL Divergence, for which $\varphi(t) = t\log t$,
    \item The Total Variation Distance, denoted as $\|\nu-\mu\|_{\rm TV}$, for which $\varphi(t) = |t-1|$,
    \item The Hellinger Divergence of order $\alpha\geq 0$, denoted as $\calH_\alpha(\nu\|\mu)$, for which $\varphi(t) = \frac{1}{\alpha-1}(t^\alpha-1)$. The case $\alpha=2$ corresponds to the $\chi^2$-Divergence, denoted as $\chi^2(\nu\|\mu)$.
\end{itemize}
All statistical distances defined above satisfy the Data-Processing Inequality (DPI), which for a divergence $D(\cdot\|\cdot)$ and a Markov kernel $K$ reads
\begin{equation*}
    D(\nu K\|\mu K) \leq D(\nu\|\mu)
\end{equation*}
for all $\nu, \mu\in\calP(\X)$ with $\nu\ll\mu$. Typically, the inequality is strict and one can give a more quantitative statement. For a fixed pair $(\mu, K)$, the DPI can be tightened by considering the associated distribution-dependent SDPI constant
\begin{equation}\label{eq:dd_sdpi_constant}
    \eta_{D}(\mu, K) = \sup_{\substack{\nu\in\calP(\X): \\0<D(\nu\|\mu)<\infty}} \frac{D(\nu K\|\mu K)}{D(\nu\|\mu)},
\end{equation}
yielding the stronger inequality ${D(\nu K\|\mu K) \leq \eta_{D}(\mu, K)D(\nu\|\mu)}$ for all $\nu\in\calP(\X)$. One can similarly define a quantity which only depends on $K$, leading to the distribution-independent SDPI constant
\begin{equation}\label{eq:di_sdpi_constant}
    \eta_{D}(K) = \sup_{\mu\in\calP(\X)}\eta_D(\mu, K) = \sup_{\substack{\nu, \mu\in\calP(\X)\\0<D(\nu\|\mu)<\infty}} \frac{D(\nu K\|\mu K)}{D(\nu\|\mu)}.
\end{equation}
In the rest of the article, we use the following notation to refer to SDPI constants under various common divergences: $\eta_{\rm TV}$ for Total Variation Distance, $\eta_{\chi^2}$ for $\chi^2$-Divergence, $\eta_{\calH_\alpha}$ for Hellinger Divergence of order $\alpha$ and $\eta_\alpha$ for R\'enyi Divergence of order $\alpha$.

\section{Main Results}\label{section:main_results}\noindent
The upcoming sections establish a variety of bounds and identities for the SDPI constants associated with R\'enyi Divergences. All the proofs can be found in the appendix.

\subsection{Bounds on the R\'enyi-SDPI Constants}\noindent
In the study of contraction coefficients for $\varphi$-Divergences, certain bounds appear quite extensively in the literature, such as the lower bounds $\eta_{\varphi}(\mu, K)\geq\eta_{\chi^2}(\mu, K)$ and $\eta_{\varphi}(K)\geq\eta_{\chi^2}(K)$, which hold for all thrice differentiable $\varphi$ with $\varphi^{\prime\prime}(1)>0$~\cite[Theorem III.3]{raginsky2016strong}. While the same results have been shown to also hold when considering the contraction of R\'enyi Divergence~\cite{jin2024properties}, it is not known whether better lower bounds exist. Regarding upper bounds, the classical result $\eta_\varphi(K) \leq\eta_{\rm TV}(K)$~\cite[Theorem III.1]{raginsky2016strong}, which holds for all $\varphi$-Divergence, is known not to extend to the R\'enyi Divergence~\cite[Example 1]{esposito2024lower}. As we will see, certain contraction properties of $\varphi$-Divergences are mirrored whenever $\alpha\in[0,1]$, whereas the case $\alpha>1$ exhibits distinct behaviours.

\begin{figure}
    \centering
    \includegraphics[width=\linewidth]{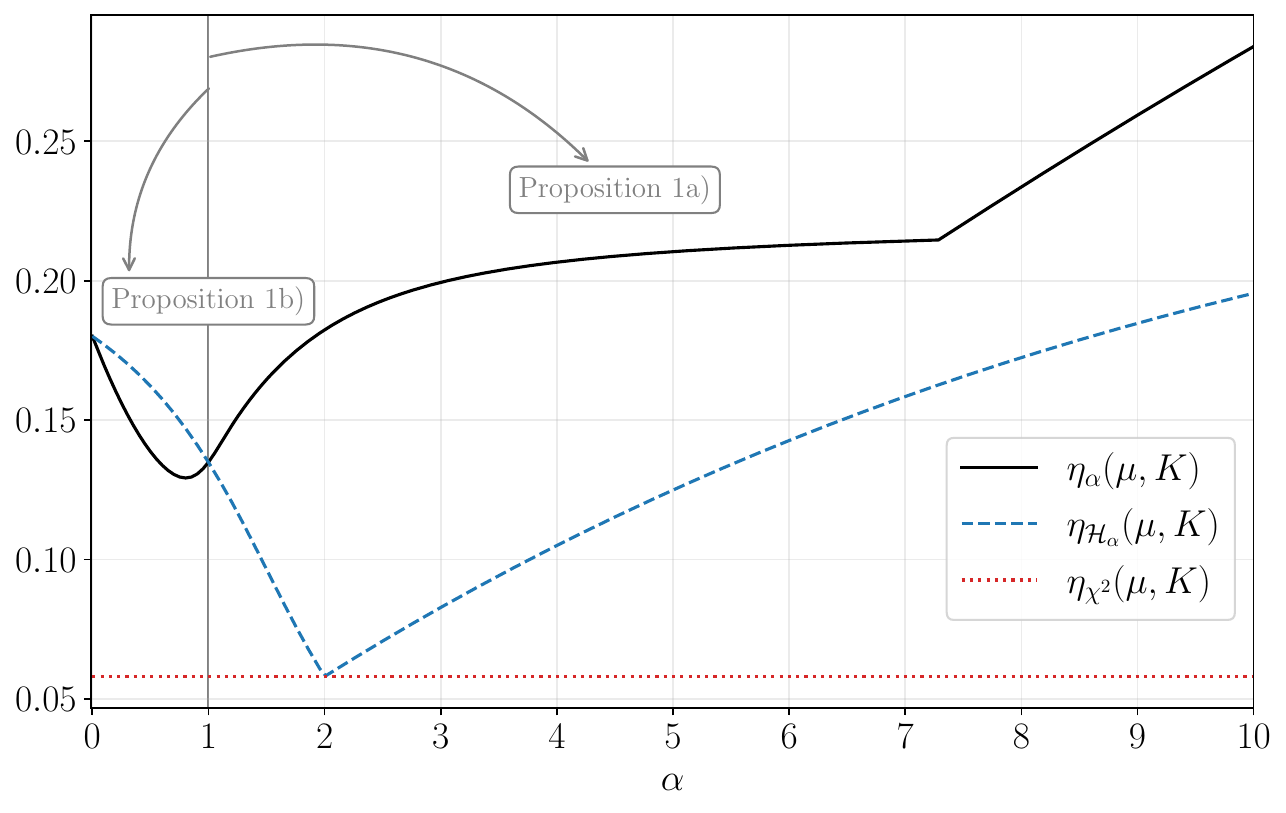}
    \caption{Comparison between the SDPI constants $\eta_{\alpha}(\mu, K)$, $\eta_{\calH_\alpha}(\mu, K)$ and $\eta_{\chi^2}(\mu, K)$ for $\alpha\in(0, 10]$, where $K=\begin{bmatrix}0.5 & 0.5\\ 0.1 & 0.9\end{bmatrix}$ and $\mu = \begin{bmatrix}0.9 & 0.1\end{bmatrix}$. The two regimes outlined in~\cref{corollary:sdpi_ub_lb_hellinger} can be observed depending on the range of $\alpha$.}
    \label{fig:comparison_chi_squared_lb_vs_hellinger_lb}
\end{figure}

Our first step is to refine known bounds on the R\'enyi-SDPI constant by relating it to the SDPI constant associated with Hellinger Divergences. This connection is made possible through the use of~\cref{lemma:sdpi_relationship_concave_function} (see~\cref{appendix:sdpi_relationship_concave_function}), a general result to compare the SDPI constants of divergences which are in one-to-one relationship, alongside the identity ${\renyiDiv{\nu}{\mu}=\frac1{\alpha-1}\log\big(1+(\alpha-1)\calH_\alpha(\nu\|\mu)\big)}$. The following result will be crucial in subsequent sections for establishing similarities and differences between the contraction of R\'enyi Divergences and $\varphi$-Divergences.
\begin{proposition}\label{corollary:sdpi_ub_lb_hellinger}
    For any pair $(\mu, K)\in\calP(\X)\times\calP(\Y|\X)$:
    \begin{enumerate}[a)]
        \item If $\alpha>1$, then
        \begin{equation*}
            \eta_\alpha(\mu, K) \geq \eta_{\calH_\alpha}(\mu, K)\quad\text{and}\quad\eta_\alpha(K) \geq \eta_{\calH_\alpha}(K)
        \end{equation*}
        \item If $\alpha\in[0,1)$, then
        \begin{equation*}
            \eta_\alpha(\mu, K) \leq \eta_{\calH_\alpha}(\mu, K)\quad\text{and}\quad\eta_\alpha(K) \leq \eta_{\calH_\alpha}(K)
        \end{equation*}
    \end{enumerate}
\end{proposition}
\begin{remark}
    Both the R\'enyi and Hellinger Divergence correspond to the KL Divergence when $\alpha\to1$, so that their SDPI constants become equal in the limit.
\end{remark}
While Part a) of~\cref{corollary:sdpi_ub_lb_hellinger} specifies new lower bounds on the R\'enyi-SDPI constant, a natural question is whether they improve over those involving $\eta_{\chi^2}$. Such an improvement can be expected since in general $\eta_{\chi^2}(\mu, K) \leq \eta_{\calH_\alpha}(\mu, K)$, and it can be observed in~\cref{fig:comparison_chi_squared_lb_vs_hellinger_lb} in the regime $\alpha>1$ where $\eta_{\calH_\alpha}(\mu, K)$ acts as a lower bound.
\subsection{Similarities and Differences with Contraction of \mbox{$\varphi$-Divergences}}\noindent
We first demonstrate that similarly to $\varphi$-Divergences~\cite{ordentlich2021strong}, the maximisation in~\cref{eq:di_sdpi_constant} can be restricted to pairs of binary supported probability measures when considering R\'enyi Divergences.
\begin{theorem}\label{prop:eta_parameter_renyi_binary}
    Let $K\in\calP(\Y|\X)$ be a Markov kernel and $\alpha\in[1, \infty]$. Then, the supremum in~\cref{eq:di_sdpi_constant} characterising $\eta_\alpha(K)$ is achieved by a pair of probability measures that are supported on a common set of at most two points in $\X$.
\end{theorem}
Our next result establishes when the Dobrushin coefficient $\eta_{\rm TV}(K)$ acts as an upper bound on $\eta_\alpha(K)$. This is made precise in the statement below, which delineates the range of $\alpha$ where the Dobrushin coefficient upper bound is valid, as well as the region where it is violated. 
\begin{proposition}\label{prop:bound_on_distribution_independent_sdpi}
    Consider $\alpha\geq 0$. Then:
    \begin{enumerate}[a)]
        \item If $\alpha\in[0,1]$, $\eta_{\alpha}(K) \leq \eta_{\rm TV}(K)$ for any Markov kernel $K\in\calP(\Y|\X)$,
        \item If $\alpha>1$, there exists a Markov kernel $K\in\calP(\Y|\X)$ such that $\eta_{\alpha}(K) > \eta_{\rm TV}(K)$.
    \end{enumerate}
\end{proposition}
Next, we identify the conditions under which R\'enyi-SDPI constants fail to guarantee a non-trivial contraction. For $\varphi$-Divergence, the condition $\eta_\varphi(K)=1$ is known to be equivalent to the existence of orthogonal rows in $K$ (when seen as a stochastic matrix)~\cite{cohen1993relative}. For R\'enyi Divergences, the same condition has been shown to imply $\eta_\alpha(K)=1$~\cite[Proposition 3]{grosse2025bounds}. We strengthen it below by providing an equivalence between structural properties of $K$ and the fact $\eta_\alpha(K)=1$ depending on the range of $\alpha$.
\begin{theorem}\label{prop:non_trivial_sdpi_characterisation}
    Consider a Markov kernel $K\in\calP(\Y|\X)$. Then we have:
    \begin{enumerate}[a)]
        \item For $\alpha\in[0,1]$, $\eta_\alpha(K)=1$ if and only if there exist  $x, x^\prime\in\X$ such that ${\supp\big(K(\cdot|x)\big)\cap\supp\big(K(\cdot|x^\prime)\big)=\emptyset}$,
        \item For $\alpha>1$, $\eta_\alpha(K)=1$ if and only if there exist  ${x, x^\prime\in\X}$ such that ${\supp\big(K(\cdot|x)\big)\neq\supp\big(K(\cdot|x^\prime)\big)}$.
    \end{enumerate}
\end{theorem}
We now provide an example of Markov kernel that illustrates both Part b) of~\cref{prop:bound_on_distribution_independent_sdpi} and Part b) of~\cref{prop:non_trivial_sdpi_characterisation}.
\begin{example}
    Consider the Z-channel $K = \begin{bmatrix}
        1 & 0\\
        1-\lambda & \lambda
    \end{bmatrix}$ with $\lambda\in(0,1)$. Then $\eta_\alpha(K)=1$ for $\alpha>1$ because the conditionals have non-identical support and moreover in that case ${\eta_{\rm TV}(K)=\lambda<1=\eta_\alpha(K)}$.
\end{example}
\Cref{prop:bound_on_distribution_independent_sdpi} and ~\cref{prop:non_trivial_sdpi_characterisation} indicate that there is a clear split depending on the order $\alpha$. To further prove this point, let us state an additional result for $\alpha\in[0,1]$ that highlights another similarity to $\varphi$-Divergences.
\begin{theorem}\label{prop:distribution_independent_sdpi_equal_chi_square_sdpi}
    Let $K$ be any Markov kernel and let $\alpha\in[0,1]$. Then, we have
    \begin{equation*}
        \eta_\alpha(K) = \eta_{\chi^2}(K).
    \end{equation*}
\end{theorem}
For $\varphi$-Divergences, the identity ${\eta_\varphi(K)=\eta_{\chi^2}(K)}$ is known to hold for all operator convex $\varphi$~\cite[Corollary III.1]{raginsky2016strong}. Once again, such an identity fails for $\alpha>1$, as exemplified by the choice $\alpha=2$ and $K=\mathrm{BSC}(\varepsilon)$ with $\varepsilon\in(0,1)$ for which $\eta_2(K)\neq\eta_{\chi^2}(K)$~\cite[Corollary 6]{jin2024properties}\footnote{We note that~\cite[Corollary 6]{jin2024properties} erroneously claimed that for ${K=\mathrm{BSC}(\varepsilon)}$ the uniform distribution $\mu_{\rm uniform}$ would be the maximiser in $\sup_\mu\eta_2(\mu, K)=\eta_2(K)$. Nevertheless, it still holds that $\eta_2(K)\neq\eta_{\chi^2}(K)$ since ${\eta_2(K)\geq \eta_2\left(\mu_{\rm uniform}, K\right)>\eta_{\chi^2}(K)}$.}.
\subsection{Contraction Properties for $\alpha\to\infty$}\noindent
The previous section demonstrates that there is a distinction depending on the range of $\alpha$; for $\alpha\in(0,1)$, the SDPI constant is akin to that of $\varphi$-Divergences, whereas for $\alpha>1$ there are discrepancies. In this section, we unravel some of the properties of the SDPI constants in the latter case, specifically for the order $\alpha\to\infty$.

An extension of~\cite[Theorem 4]{jin2024properties}---which states that the distribution achieving the supremum in the definition of $\eta_2(\mu, K)$ has support strictly smaller than the one of $\mu$---can be obtained for $\alpha\to\infty$ (see~\cref{appendix:proof_boundary_nu}). In reality, the following result shows that the achieving distribution $\nu$ is not merely of smaller support, but is precisely the restriction of $\mu$ to a subset $A\subsetneq\X$.
\vspace{1.5em}
\begin{theorem}\label{prop:infty_sdpi_achieved_on_conditionals}
    Consider any pair $(\mu, K)\in\calP(\X)\times\calP(\Y|\X)$. Then
    \begin{equation*}
        \eta_\infty(\mu, K) = \sup_{A\subsetneq\X}\frac{D_\infty\left(\mu_{|A} K\|\mu K\right)}{D_\infty\left(\mu_{|A}\|\mu\right)},
    \end{equation*}
    where $\mu_{|A}(x) \triangleq \frac{\mathbbm{1}_A(x)\mu(x)}{\mu(A)}$ denotes the restriction of $\mu$ to the event $A$.
\end{theorem}
Even with the simpler expression put forth in~\cref{prop:infty_sdpi_achieved_on_conditionals}, the computation of $\eta_\infty(\mu, K)$ can still be prohibitive. This can be addressed by bounding this quantity with the SDPI constant of a different divergence that is more amenable to analytical evaluation. The following result establishes bounds on $\eta_\infty(\mu, K)$ via the contraction of Total Variation Distance.
\begin{proposition}\label{prop:bounds_on_renyi_infty_sdpi_by_tv}
    Consider a pair $(\mu, K)\in\calP(\X)\times\calP(\Y|\X)$. Then we have
    \begin{equation}\label{eq:ub_on_renyi_infty_sdpi_by_tv}
    \eta_\infty(\mu, K) \leq \frac{\eta_{\rm TV}(\mu, K)}{\min_{y\in\supp(\mu K)}\mu K(y)}.
    \end{equation}
    Moreover, if $\mu$ has full-support,
    \begin{equation}\label{eq:lb_on_renyi_infty_sdpi_by_tv}
        \eta_\infty(\mu, K)\geq\eta_{\rm TV}(\mu, K)\cdot\min_{x\in\X}\mu(x).
    \end{equation}
\end{proposition}
The presence of minima in~\cref{prop:bounds_on_renyi_infty_sdpi_by_tv} is reminiscent of certain Pinsker-type inequalities involving R\'enyi Divergence~\cite{sason2015reverse}. This is, however, not a coincidence: such inequalities form an effective way to compare the contraction of different $\varphi$-Divergences~\cite{makur2020comparison, george2025rate,grosse2025bounds}.
\begin{figure}
    \centering
    \includegraphics[width=\linewidth]{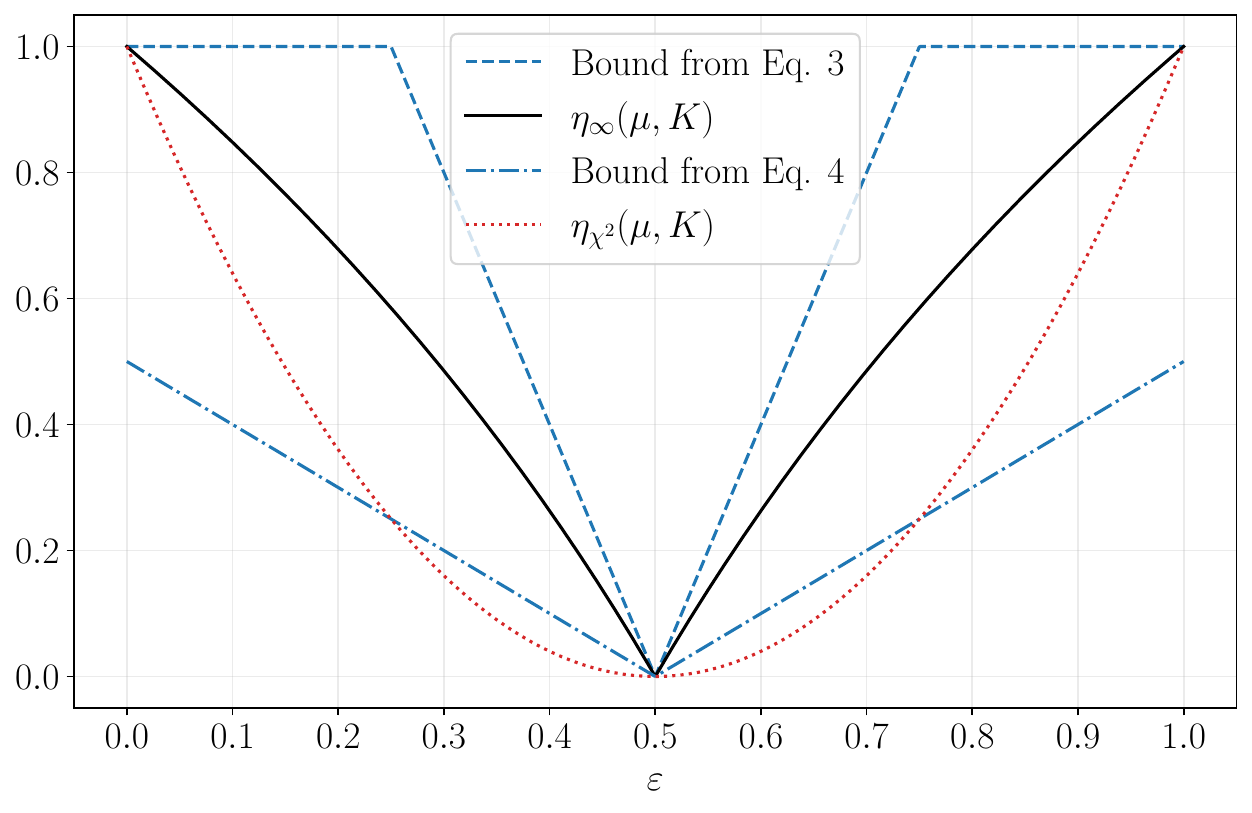}
    \caption{Plot of the bounds in~\cref{prop:bounds_on_renyi_infty_sdpi_by_tv} for $K=\mathrm{BSC(\varepsilon)}$ and $\mu=\begin{bmatrix}
        0.5&0.5
    \end{bmatrix}$ together with $\eta_{\chi^2}(\mu, K)$ for comparison.}
    \label{fig:bounds_on_renyi_infty_sdpi_by_tv}
\end{figure}

\Cref{fig:bounds_on_renyi_infty_sdpi_by_tv} displays the bounds in~\cref{prop:bounds_on_renyi_infty_sdpi_by_tv} for the choice $K=\mathrm{BSC}(\varepsilon)$ and $\mu = \begin{bmatrix}
        0.5&0.5
    \end{bmatrix}$. In this setting, $\eta_\infty(\mu, K)$ can be determined via~\cref{prop:infty_sdpi_achieved_on_conditionals} as ${\eta_\infty(\mu, K) = 1-\log_2\frac1{\max
    (\varepsilon, 1-\varepsilon)}}$ and ${\eta_{\rm TV}(\mu, K)=\eta_{\rm TV}(K)=|1-2\varepsilon|}$. The upper bound from~\cref{eq:ub_on_renyi_infty_sdpi_by_tv} gives a non trivial result when $\varepsilon\in(0.27, 0.75)$. The lower bound from~\cref{eq:lb_on_renyi_infty_sdpi_by_tv}, on the other hand, is always non-trivial and offers an improvement over the local bound $\eta_\infty(\mu, K)\geq \eta_{\chi^2}(\mu, K)=(1-2\varepsilon)^2$ when $\varepsilon\in(0.27, 0.75)$. Both bounds become tight as $\varepsilon\to0.5$.

Equipped with~\cref{prop:infty_sdpi_achieved_on_conditionals}, a characterisation of the distribution-independent SDPI constant associated with the $\infty$-R\'enyi Divergence can be uncovered.
\begin{theorem}\label{prop:infty_renyi_sdpi_closed_form}
    For any Markov kernel $K\in\calP(\Y|\X)$, the distribution-independent SDPI constant associated with the R\'enyi Divergence of order $\infty$ can be written as
    \begin{equation*}
        \eta_\infty(K) = \sup_{x, x^\prime\in\X}\left\{1 -\min_{y\in\supp(K(\cdot|x))} \frac{K(y|x^\prime)}{K(y|x)}\right\}.
    \end{equation*}
\end{theorem}
\begin{figure}
    \centering
    \includegraphics[width=\linewidth]{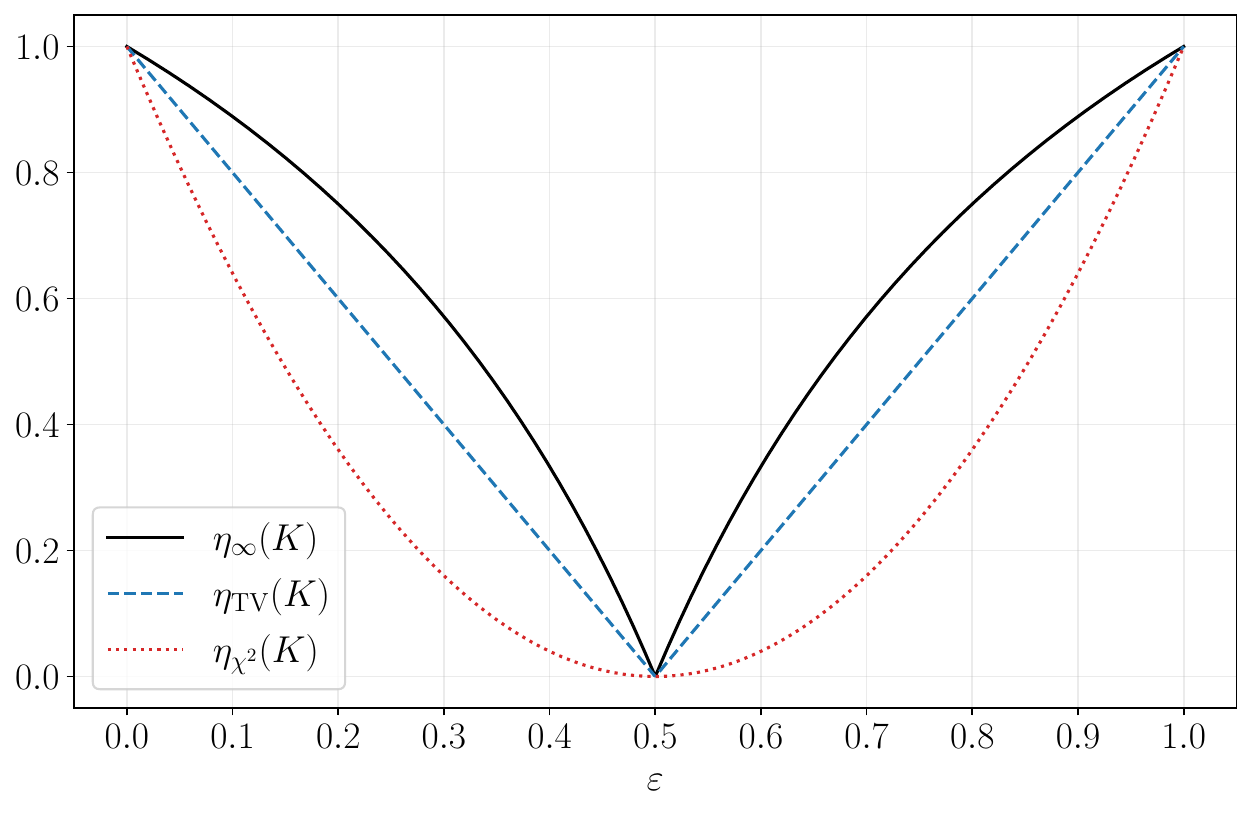}
    \caption{Comparison of distribution-independent SDPI constants for $K=\mathrm{BSC(\varepsilon)}$.}
    \label{fig:bsc_sdpi_constants_comparison}
\end{figure}
The following example shows an application of~\cref{prop:infty_renyi_sdpi_closed_form} for the Binary Symmetric Channel.
\begin{example}
    For the choice $K=\mathrm{BSC}(\varepsilon)$, one obtains $\eta_\infty(K) = 1 - \min\left(\frac{\varepsilon}{1-\varepsilon}, \frac{1-\varepsilon}{\varepsilon}\right)$. A comparison with other common SDPI constants in this setting is displayed in~\cref{fig:bsc_sdpi_constants_comparison}.
\end{example}
To close this section, let us make through~\cref{prop:infty_renyi_sdpi_closed_form} a direct connection with the condition known as ultra-mixing in the analysis of Markov chains contraction~\cite{del2003contraction}.
\begin{corollary}\label{corollary:equivalence_ultra_mixing}
    Let $\varepsilon\in[0,1]$. Then ${\eta_\infty(K)\leq 1-\varepsilon}$ if and only if $\varepsilon$-ultra-mixing holds, that is $\frac{K(y|x^\prime)}{K(y|x)} \geq\varepsilon$ for all $x, x^\prime\in\X$ and ${y\in\supp(K(\cdot|x))}$.
\end{corollary}
\section{Connections and Applications}\label{section:applications}\noindent
\subsection{Relation to $\varepsilon$-Local Differential Privacy}\noindent
In this section, we show how the concept of $\varepsilon$-Local Differential Privacy (LDP) is equivalent to the contraction of R\'enyi Divergence of order $\infty$. A similar connection has been drawn between $(\varepsilon, \delta)$-LDP and the contraction of the Hockey-Stick Divergence~\cite{asoodeh2020contraction}. Let us recall that a privacy mechanism is a (random) mapping from elements of a set $\X$ to elements of a set $\Y$, which is usually represented by some Markov kernel $K\in\calP(\Y|\X)$. In the following, we shall only focus on $\varepsilon$-LDP, described below.
\begin{definition}
    Let $\varepsilon\geq 0$. A privacy mechanism $K\in\calP(\Y|\X)$ is $\varepsilon$-LDP if
    \begin{equation*}
        \sup_{x, x^\prime\in\X}\sup_{A\subseteq\X}\frac{K(A|x^\prime)}{K(A|x)}\leq e^\varepsilon.
    \end{equation*}
\end{definition}
The relationship with the SDPI constant of $\infty$-R\'enyi Divergence can then be formulated via~\cref{prop:infty_renyi_sdpi_closed_form}.
\begin{corollary}
    A privacy mechanism $K\in\calP(\Y|\X)$ is $\varepsilon$-LDP if and only if $\eta_\infty(K)\leq 1-e^{-\varepsilon}$.
\end{corollary}
Note that~\Cref{corollary:equivalence_ultra_mixing} also allows us to view ultra-mixing as being equivalent to $\varepsilon$-LDP, a link that was previously observed in~\cite[Table 1]{balle2019privacy}.
\subsection{Mixing of Markov Chains}\label{section:mixing_time}\noindent
SDPIs provide a powerful framework to analyse the convergence of Markov chains, a topic of great practical importance notably in the context of Markov chain Monte-Carlo algorithms~\cite{levin2017markov,montenegro2006mathematical,gilks1995markov}. Specifically, a Markov kernel $K\in\calP(\X|\X)$ induces a Markov chain, and one can study the $n$-step evolution of any distribution $\nu$ by looking at $\nu K^n$. When a Markov chain is irreducible and aperiodic, it is well-known that it admits a unique stationary distribution $\pi$ (a distribution such that $\pi=\pi K$), and $\nu K^n$ converges to $\pi$ for any $\nu$ as a function of $n$.

Traditionally, convergence is quantified in terms of \mbox{$L^\alpha$-norms} for $\alpha\geq 1$ via bounds~\cite{rudolf2012explicit} such as
\begin{equation}\label{eq:mcmt_linear_sdpi}
    \left\|\frac{\dd\nu K^n}{\dd\pi} - 1\right\|_{L^\alpha(\pi)} \leq \gamma_\alpha^n\left\|\frac{\dd\nu}{\dd\pi} - 1\right\|_{L^\alpha(\pi)} \text{ for } \gamma_\alpha\in[0,1],
\end{equation}
indicating that $\left\|\frac{\dd\nu K^n}{\dd\pi} - 1\right\|_{L^\alpha(\pi)}$ converges to 0 at geometric rate given by $\gamma_\alpha$. In contrast, using the R\'enyi-SDPI gives ${\renyiDiv{\nu K^n}{\pi} \leq \eta_\alpha(\pi, K)^n \renyiDiv{\nu}{\pi}}$, which can be equivalently formulated as follows.
\begin{proposition}\label{prop:mcmt}
    Consider a Markov kernel $K\in\calP(\X|\X)$ whose induced Markov chain is aperiodic and irreducible, with stationary distribution $\pi\in\calP(\X)$. Then for any $\alpha>1$ and $n\geq 1$, one has
    \begin{equation}\label{eq:mcmt_non_linear_sdpi}
        \left\|\frac{\dd\nu K^n}{\dd\pi}\right\|_{L^\alpha(\pi)} \leq \left\|\frac{\dd\nu}{\dd\pi}\right\|_{L^\alpha(\pi)}^{\eta_\alpha(\pi, K)^n}
    \end{equation}
    for all $\nu\in\calP(\X)$.
\end{proposition}
This type of contraction has already appeared in the literature, and is referred to as ``non-linear''~\cite{polyanskiy2017strong,du2017strong,gu2023non}. In this form, it still quantifies the convergence of $\nu K^n$ to $\pi$, but this time through the convergence of $\left\|\frac{\dd\nu K^n}{\dd\pi}\right\|_{L^\alpha(\pi)}$ to 1.

The two forms of convergence are not immediately comparable in general. However, for $\alpha=2$, \cref{eq:mcmt_linear_sdpi} can be written as
\begin{equation}\label{eq:mcmt_linear_sdpi_alpha_2}
    \left\|\frac{\dd\nu K^n}{\dd\pi} - 1\right\|_{L^2(\pi)}^2 \leq \eta_{\chi^2}(\pi, K)^n\left\|\frac{\dd\nu}{\dd\pi} - 1\right\|_{L^2(\pi)}^2,
\end{equation}
while~\cref{eq:mcmt_non_linear_sdpi} reads
\begin{equation}\label{eq:mcmt_non_linear_sdpi_alpha_2}
    \left\|\frac{\dd\nu K^n}{\dd\pi}-1\right\|_{L^2(\pi)}^2 \leq \left(\left\|\frac{\dd\nu}{\dd\pi}\right\|_{L^2(\pi)}^{2}\right)^{\eta_2(\pi, K)^n}-1,
\end{equation}
thus enabling a direct comparison.
\begin{figure}
    \centering
    \includegraphics[width=\linewidth]{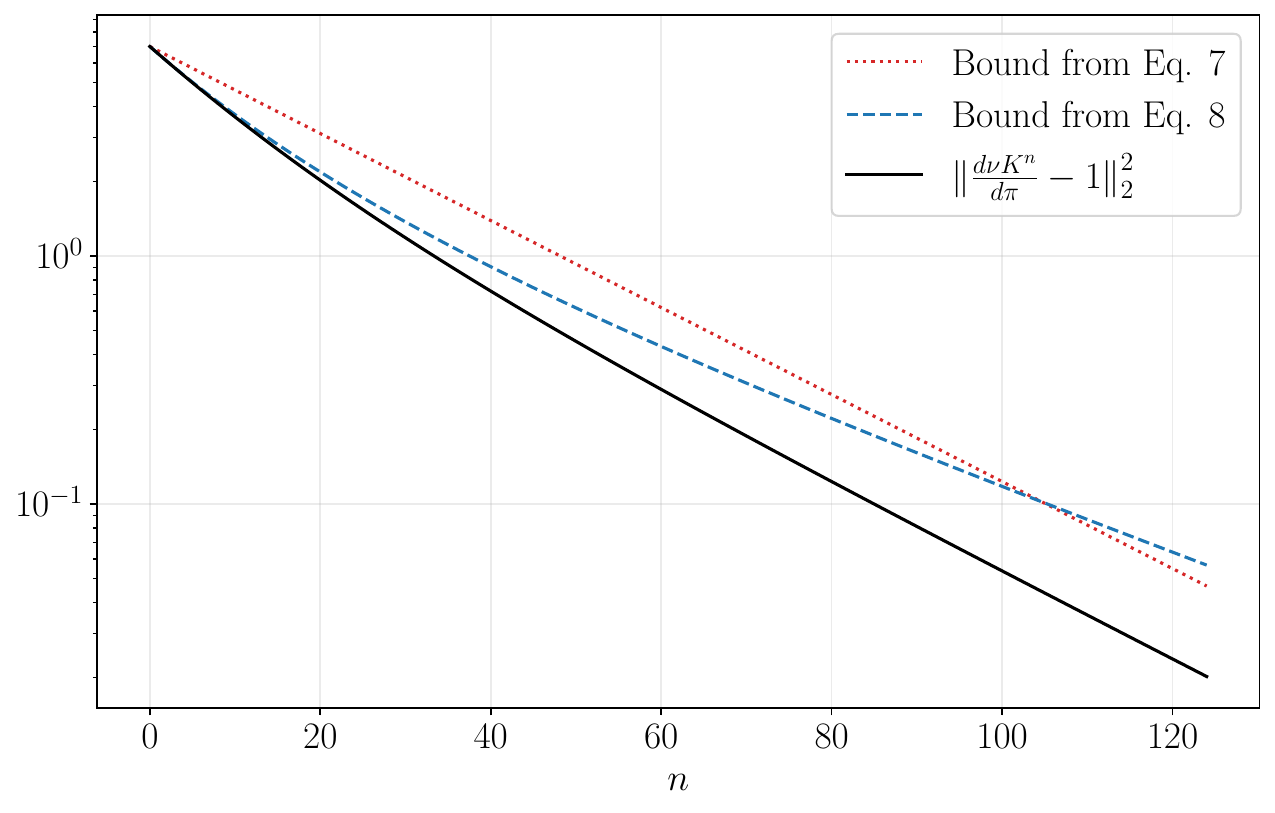}
    \caption{Plot of the bounds on $\left\|\frac{\dd\nu K^n}{\dd\pi}-1\right\|_{L^2(\pi)}^2$ from~\cref{eq:mcmt_non_linear_sdpi_alpha_2,eq:mcmt_linear_sdpi_alpha_2} as a function of $n$. The Markov chain corresponds to ${K=\mathrm{BSC}(\varepsilon)\otimes \mathrm{BSC}(\varepsilon)\otimes \mathrm{BSC}(\varepsilon)}$ with $\varepsilon=10^{-2}$. The stationary distribution $\pi$ is uniform and $\nu$ is a Dirac mass. The SDPI constants appearing in the bounds are evaluated numerically. Additional details can be found in~\cref{appendix:mixing_time_experiment}.}
    \label{fig:bsc_tensored_mixing_times}
\end{figure}

To illustrate the benefits of~\cref{eq:mcmt_non_linear_sdpi_alpha_2} over~\cref{eq:mcmt_linear_sdpi_alpha_2}, we consider the setting described in~\cref{appendix:mixing_time_experiment} with $\alpha=2$. \Cref{fig:bsc_tensored_mixing_times} displays the bounds in~\cref{eq:mcmt_linear_sdpi_alpha_2,eq:mcmt_non_linear_sdpi_alpha_2} in that scenario. One can observe that while the bound provided by the R\'enyi-SDPI may not always improve over the classical one based on $L^\alpha$-norms, it can give tighter results for small values of $n$. This in particular indicates that a reduced number of iterations may be required to reach a particular value of $\left\|\frac{\dd\nu K^n}{\dd\pi} - 1\right\|_{L^2(\pi)}^2$. We leave a more in-depth study of Markov chains convergence for future work.

\section*{Acknowledgment}\noindent
This work was supported in part by the Swiss
National Science Foundation under Grant 200364, and was partially conducted during the first author's visit at the Okinawa Institute of Science and Technology. We also thank Anuj Yadav for useful discussions about the case $\alpha\to\infty$ of~\cref{prop:eta_parameter_renyi_binary}.
\bibliographystyle{IEEEtran}
\bibliography{references}
\newpage
\onecolumn
\appendix
\crefalias{subsection}{appendix}
\subsection{Some Facts about Markov Kernels}\noindent
    This section collects some basic facts about Markov Kernels that are used in the proofs.

    While~\cref{sec:markov_kernels} introduces Markov kernels as acting on probability measures, one can also view them as acting on functions $f\in\F(\Y)$ as
    \begin{equation*}
        Kf(x) = \sum_{y\in\Y} f(y)K(y|x), \quad \text{for }x\in\X.
    \end{equation*}
    For any pair $(\mu, K)\in\calP(\X)\times\calP(\Y|\X)$, there exists a unique (dual) kernel $\dualK\in\calP(\X|\Y)$ such that
    \begin{equation*}
        \ip{Kf}{g}_{\mu} = \ip{f}{\dualK g}_{\mu K}, \quad\forall f\in\F(\Y), g\in\F(\X) 
    \end{equation*}
    where we have used the inner product notation ${\ip{f}{g}_\mu\triangleq \mean{\mu}{f\cdot g}=\sum_{x\in\X}f(x)g(x)\mu(x)}$. In discrete settings like ours, the transition probabilities of this dual kernel can be explicitly computed as
    \begin{equation*}
        \dualK(x|y) = \frac{\mu(x)K(y|x)}{\mu K (y)}.
    \end{equation*}
    We also note that for a pair $(\mu, K)$ and any other probability measure $\nu\ll\mu$, we have
    \begin{equation}\label{eq:dual_kernel_formula}
        \dualK f = \frac{\dd(\nu K)}{\dd(\mu K)}\quad \mu K-\text{a.e.},
    \end{equation}
    where $f=\frac{\dd\nu}{\dd\mu}$~\cite[Lemma 1]{esposito2024contraction}.
\subsection{Relating SDPI Constants of Divergences in One-to-One Correspondence}\label{appendix:sdpi_relationship_concave_function}\noindent
    \begin{lemma}\label{lemma:sdpi_relationship_concave_function}
        Consider any divergence ${D(\cdot\|\cdot):\calP(\X)\times\calP(\X)\to\reals_+}$ that satisfies the DPI. Let ${g:\reals_+\to\reals_+}$ be a function such that $g(0)=0$, and let $D^{(g)}(\nu\|\mu)=g\big(D(\nu\|\mu)\big)$. Then, for any admissible pair $(\mu, K)$:
        \begin{enumerate}[a)]
            \item If $g$ is convex, then
            \begin{equation*}
                \eta_{D^{(g)}}(\mu, K) \leq \eta_{D}(\mu, K)\quad\text{and}\quad    \eta_{D^{(g)}}(K) \leq \eta_{D}(K),
            \end{equation*}
            
            \item If $g$ is concave, then
            \begin{equation*}
                \eta_{D^{(g)}}(\mu, K) \geq \eta_{D}(\mu, K)\quad\text{and}\quad  \eta_{D^{(g)}}(K) \geq \eta_{D}(K).
            \end{equation*}
        \end{enumerate}
    \end{lemma}
    \begin{IEEEproof}
        Let us consider the case where $g$ is convex first. Since $g$ is non-negative and $g(0)=0$, $g$ must be non-decreasing so that $D^{(g)}$ also satisfies the DPI. Moreover in this case, the function $x\mapsto\frac{g(x)}{x}$ is non-decreasing for $x>0$. Indeed, by convexity of $g$ and the fact that $g(0)=0$, we have $g\big(\lambda x+(1-\lambda)\cdot0\big)\leq \lambda g(x)$ for $\lambda\in[0,1]$. Letting $y=\lambda x\in[0, x]$, the inequality reads $\frac{g(y)}{y}\leq\frac{g(x)}{x}$, establishing non-decreasibility. Choosing $y=D(\nu K\|\mu K) \leq D(\nu\|\mu)=x$ in the last inequality and re-arranging gives
        \begin{equation*}
            \frac{D^{(g)}(\nu K\|\mu K)}{D^{(g)}(\nu\|\mu)}\leq\frac{D(\nu K\|\mu K)}{D(\nu\|\mu)}.
        \end{equation*}
        Taking supremum over $\nu$ on each side gives $\eta_{D^{(g)}}(\mu, K) \leq \eta_{D}(\mu, K)$ and further taking supremum over $\mu$ gives the distribution-independent result. The same argument applies when $g$ is concave, but with reversed inequalities.
    \end{IEEEproof}
\subsection{Proof of~\cref{prop:eta_parameter_renyi_binary}}\label{appendix:proof_eta_parameter_renyi_binary}\noindent
    We closely follow the proof in~\cite{ordentlich2021strong}, where the case $\alpha\to1$ was proven, and adapt it where necessary. 
    First, we consider $\alpha>1$. For two distributions $\nu, \mu \in \mathcal{P}(\X)$ and $\lambda\in(0,1)$, define
    \begin{equation*}
        M_{\lambda}(\nu, \mu) \triangleq H_{\alpha}(\nu K \| \mu K) - \left[H_{\alpha}(\nu \| \mu)\right]^{\lambda}
    \end{equation*}
    where $H_\alpha(\nu \| \mu) \triangleq \mean{\mu}{\left(\frac{\dd \nu}{\dd \mu}\right)^{\alpha}}$. This quantity is relevant because for any $\lambda\in(0,1)$ such that  $\eta_\alpha(K)\geq \lambda$, this equivalently means there must exist a pair $(\nu, \mu)$ with $M_\lambda(\nu, \mu)\geq 0$, and hence
    \begin{equation*}
        \eta_{\alpha}(K) = \sup\left\{\lambda\in(0,1): \sup_{\nu, \mu} M_\lambda(\nu, \mu)\geq 0\right\}.
    \end{equation*}
    We will now show that the inner supremum $\sup_{\nu, \mu} M_\lambda(\nu, \mu)$ can be restricted to distributions supported on at most two points.

    Fix any $(\nu, \mu)$ and let $r(x)=\frac{\nu(x)}{\mu(x)}$. We are searching for a distribution $\hat{\mu}$ that can potentially increase the value of $M_\lambda(\nu, \mu)$. Once $\hat{\mu}$ is fixed, we associate to it the distribution $\hat{\nu}$ through $\hat{\nu}(x)=r(x)\hat{\mu}(x)$. Since both $\hat{\nu}, \hat{\mu}$ must be valid probability distributions, we must ensure $\sum_x\hat{\mu}(x)=1$ and $\sum_x\hat{\nu}(x)=\sum_x r(x)\hat{\mu}(x)=1$. This defines the space of potential $\hat{\mu}$ to consider as
    \begin{equation}
        \calS\triangleq \bigg\{\hat{\mu}\in\mathcal{P}\big(\mathrm{supp}(\mu)\big): \sum_{x\in\X} \frac{\nu(x)}{\mu(x)}\cdot \hat{\mu}(x)=1 \bigg\}.\nonumber
    \end{equation}
    With this choice, notice how since $\mu\in\calS$, we have that $\max_{\hat{\mu}\in\calS} M_{\lambda}(\hat{\nu},\hat{\mu}) = \max_{\hat{\mu}\in\calS} M_{\lambda}\left(\frac{\nu}{\mu}\hat{\mu},\hat{\mu}\right)\geq g(\mu) = M_{\lambda}(\nu,\mu)$.

    Next, let $g:\calS\to\reals$ be defined as $g(\hat{\mu})=M_\lambda\left(\frac{\nu}{\mu}\hat{\mu},\hat{\mu}\right)= \hellingerInt{(\frac{\nu}{\mu}\hat{\mu}) K}{\hat{\mu} K} - \left[\hellingerInt{\frac{\nu}{\mu}\hat{\mu}}{\hat{\mu}}\right]^\lambda$, which is the function we now want to maximise now that we have established it is sufficient to optimise over $\mathcal{S}$. The function $g$ is convex because 
    \begin{itemize}
        \item $\hat{\mu}\mapsto\hellingerInt{(\frac{\nu}{\mu}\hat{\mu}) K}{\hat{\mu} K}$ is convex by joint convexity of $(\nu,\mu)\mapsto \hellingerInt{\nu}{\mu}$ and since $\hat{\mu}\mapsto \big((\frac{\nu}{\mu}\hat{\mu})K, \hat{\mu}K\big)$ is linear,
        
        \item $\hat{\mu}\mapsto \left[\hellingerInt{\frac{\nu}{\mu}\hat{\mu}}{\hat{\mu}}\right]^\lambda=\left(\sum_x \hat \mu(x) \left(\frac{\nu(x)}{\mu(x)}\right)^{\alpha}\right)^\lambda\nonumber$
        is concave since it is the composition of a linear function with the concave map $x\mapsto x^\lambda$.
    \end{itemize}
    It therefore follows that $\max_{\hat{\mu}\in\calS} g(\hat{\mu})$ is obtained at an extreme point of $\calS$. Since $\calS$ is the intersection of the simplex with a hyperplane, its extreme points are supported on at most two atoms.
    
    For $\alpha\to\infty$, the approach is similar, except we directly use
    \begin{equation*}
        \eta_\infty(K)=\sup\big\{\lambda\in(0,1): \eta_\infty(K)\geq \lambda\big\}=\sup\left\{\lambda\in(0,1): \sup_{\nu, \mu} \frac{D_\infty(\nu K\|\mu K )}{D_\infty(\nu\|\mu)}\geq \lambda\right\}
    \end{equation*}
    and thus define $\widetilde{M}_\lambda(\nu, \mu) = \frac{D_\infty(\nu K\|\mu K )}{D_\infty(\nu\|\mu)}$. We use the same idea as before and define $\mathcal{S}$ similarly. The function to be maximised is now $\tilde{g}:\calS\to\reals$ defined as $\tilde{g}(\hat{\mu})=\widetilde{M}_\lambda\left(\frac{\nu}{\mu}\hat{\mu},\hat{\mu}\right) = \frac{D_\infty((\frac{\nu}{\mu}\hat{\mu}) K\|\hat{\mu} K)}{D_\infty(\frac{\nu}{\mu}\hat{\mu}\|\hat{\mu})}$. It is quasiconvex because
    \begin{itemize}
        \item $\hat{\mu}\mapsto D_\infty(\frac{\nu}{\mu}\hat{\mu}\|\hat{\mu})$ is constant since $D_\infty(\frac{\nu}{\mu}\hat{\mu}\|\hat{\mu})= \log\max_{x\in\X}\frac{\nu(x)}{\mu(x)}$.
        \item $\hat{\mu}\mapsto D_\infty((\frac{\nu}{\mu}\hat{\mu}) K\|\hat{\mu} K) = \log\sup_{y\in\Y}\frac{\sum_{x\in\X}r(x)\hat{\mu}(x)K(y|x)}{\sum_{x\in\X} \hat{\mu}(x)K(y|x)}$ is quasiconvex. Indeed, for any fixed $y$, the inner ratio is quasiconvex since it is a linear-fractional function in $\hat{\mu}$~\cite[Example 3.32]{boyd2004convex}. Moreover the supremum of quasiconvex function is quasiconvex, and further composing with the logarithm retains quasiconvexity since it is a monotonically increasing function.
    \end{itemize}
    The rest of the proof follows since a quasiconvex function is maximised at an extreme point of the space.
\subsection{Proof of~\cref{prop:bound_on_distribution_independent_sdpi}}\label{appendix:proof_bound_on_distribution_independent_sdpi}\noindent
    We first prove Part a). Note that the cases $\alpha\to0$ or $\alpha\to1$ fall back to the already-known result $\eta_{\rm KL}(K)\leq\eta_{\rm TV}(K)$. For $\alpha\in(0,1)$, \Cref{corollary:sdpi_ub_lb_hellinger} gives $\eta_\alpha(K)\leq\eta_{\calH_\alpha}(K)$ and since the Hellinger Divergence is in the family of $\varphi$-Divergences, it satisfies $\eta_{\calH_\alpha}(K)\leq\eta_{\rm TV}(K)$.
    
    Part b) is proven by considering $\X=\{0,1\}$ and the Z-channel $K = \begin{bmatrix}
        1 & 0\\
        1-\lambda & \lambda
    \end{bmatrix}$ with $\lambda\in(0, 1)$, for which $\eta_{\rm TV}(K)=\lambda$. Considering the input and output R\'enyi Divergences between the Dirac mass $\delta_1$ and $\mu_\varepsilon \triangleq \begin{bmatrix}
        1-\varepsilon&\varepsilon
    \end{bmatrix}$, we find their ratio to be
    \begin{align*}
        \frac{\renyiDiv{\delta_1 K}{\mu_\varepsilon K}}{\renyiDiv{\delta_1}{\mu_\varepsilon}} &= \frac{\frac1{\alpha-1}\log\left((1-\lambda)^\alpha(1-\lambda\varepsilon)^{1-\alpha} + \lambda^\alpha(\lambda\varepsilon)^{1-\alpha}\right)}{\log \frac1\varepsilon} \\
        &= 1 - \frac{\frac1{\alpha-1}\log\left(\lambda + (1-\lambda)^\alpha(\frac1\varepsilon-\lambda )^{1-\alpha}\right)}{\log\frac1\varepsilon}.
    \end{align*}
    Letting $u(\varepsilon) = \log\frac1\varepsilon$ and $g(x) = \frac1{\alpha-1}\log\left(\lambda + (1-\lambda)^\alpha(e^x-\lambda )^{1-\alpha}\right)$, the ratio can be expressed as $1 - \frac{g\big(u(\varepsilon)\big)}{u(\varepsilon)}$, and we find that
    \begin{equation*}
        \lim_{\varepsilon\to0}\left(1-\frac{g\big(u(\varepsilon)\big)}{u(\varepsilon)}\right) = 1 - \lim_{u\to\infty}\frac{\frac1{\alpha-1}\log\left(\lambda + \frac{(1-\lambda)^\alpha}{(e^u-\lambda )^{\alpha-1}}\right)}{u} = 1.
    \end{equation*}
    But by definition of the R\'enyi-SDPI constant, one has
    \begin{equation*}
        \lim_{\varepsilon\to0}\frac{\renyiDiv{\nu K}{\mu_\varepsilon K}}{\renyiDiv{\nu}{\mu_\varepsilon}} \leq \eta_\alpha(K)\leq 1,
    \end{equation*}
    so that $\eta_\alpha(K) = 1$ and in particular $\eta_\alpha(K) > \eta_{\rm TV}(K)$.
    
    Since this is a limiting argument, one can wonder whether there in fact exists a pair $\nu, \mu$ such that $\frac{\renyiDiv{\nu K}{\mu K}}{\renyiDiv{\nu}{\mu}}>\eta_{\rm TV}(K)$. We give a positive answer by showing that in our previous example the function $\varepsilon \mapsto 1 - \frac{g(u(\varepsilon))}{u(\varepsilon)}$ is decreasing, so that there always exists an $\varepsilon>0$ small enough such that $\frac{\renyiDiv{\delta_1 K}{\mu_\varepsilon K}}{\renyiDiv{\delta_1}{\mu_\varepsilon}} >\eta_{\rm TV}(K)$. Notice that
    \begin{equation*}
        g^\prime(x) = \frac{e^x(1-\lambda)^\alpha}{\lambda(e^x-\lambda)^\alpha + (1-\lambda)^\alpha(e^x-\lambda)} = \frac{(1-\lambda)^\alpha(1-\lambda e^{-x})^{-1}}{\lambda(e^x-\lambda)^{\alpha-1}+(1-\lambda)^\alpha},
    \end{equation*}
    and since the numerator is decreasing in $x$ and the denominator is increasing in $x$, it follows that $g^\prime(x)$ is decreasing, or in other words $g$ is strictly concave. By the same argument as in~\cref{appendix:sdpi_relationship_concave_function}, the strict concavity of $g$ together with the fact that $g(0)=0$ implies that $u\mapsto \frac{g(u)}{u}$ is decreasing. As a consequence, the map $\varepsilon\mapsto 1 - \frac{g(u(\varepsilon))}{u(\varepsilon)}$ is decreasing.
\subsection{Distribution-Dependent SDPI Constant is Achieved at the Boundary for $\alpha\in\{2, \infty\}$}\label{appendix:proof_boundary_nu}\noindent
    \begin{proposition}\label{prop:sdpi_achieved_by_boundary}
    Let $\alpha\in\{2, \infty\}$ and consider any admissible pair $(\mu, K)$. Then either
    \begin{equation*}
        \eta_\alpha(\mu, K) = \sup_{\substack{\nu\in\calP(\X):\\\supp(\nu)\subset\supp(\mu)}}\frac{\renyiDiv{\nu K}{\mu K}}{\renyiDiv{\nu}{\mu}},
    \end{equation*}
    or
    \begin{itemize}
        \item $\eta_2(\mu, K) = \eta_{\chi^2}(\mu, K)$ if $\alpha=2$,
        \item $\eta_\infty(\mu, K) = \eta_{L^\infty}(\mu, K)$ if $\alpha\to\infty$,
    \end{itemize}
    where $\eta_{L^\infty}(\mu, K)$ denotes the SDPI constant associated with the divergence $L^\infty(\nu\|\mu)\triangleq \|\frac{\dd\nu}{\dd\mu}\|_{L^\infty(\mu)}-1$.
    \end{proposition}

    \begin{IEEEproof}
    For $\alpha=2$, let us note that a proof has already been given in~\cite{jin2024properties}, but for completeness we show how our proof strategy covers this case as well. Let us rewrite the optimisation problem
    \begin{equation*}
        \eta_\alpha(\mu, K) = \sup_{\substack{\nu\in\calP(\X)\\0<\renyiDiv{\nu}{\mu}<\infty}}\frac{\renyiDiv{\nu K}{\mu K}}{\renyiDiv{\nu}{\mu}}
    \end{equation*}
    in a different manner. The space over which the optimisation is carried on is effectively ${\calP_\mu\triangleq \calP(\supp(\mu))}$. Since the probability simplex is convex and compact for finite alphabets, notice that any $\nu\in\calP_\mu$ can be written as a convex combination $(1-t)\mu+t\tilde{\nu}$ for some $t\in[0,1]$ and $\tilde{\nu}\in\partial\calP_\mu$, where $\partial\calP_\mu$ denotes the boundary of the simplex, that is the set of probability measure $\tilde{\nu}$ with $\supp(\tilde{\nu})\subset\supp(\mu)$. As a consequence, the computation of the SDPI constant can equivalently be written as
    \begin{equation*}
    \sup_{\tilde{\nu}\in\partial\calP_\mu}\sup_{t\in(0,1]}\frac{D_\alpha\big((1-t)\mu K+t\tilde{\nu}K\|\mu K\big)}{D_\alpha((1-t)\mu+t\tilde{\nu}\|\mu)}.
    \end{equation*}
    If we can show that for any $\tilde{\nu}\in\partial\calP_\mu$, the ratio of R\'enyi divergences is increasing as a function of $t$, the proof will be done since the choice $t=1$ will yield an optimisation only over the boundary of $\calP_\mu$. Equivalently, we want to show that the map
    \begin{equation}\label{eq:function_to_show_monotonicity_of}
        t\mapsto\frac{\log\mean{\mu K}{\left(1 + t\left(\frac{\dd\tilde{\nu}K}{\dd\mu K}-1\right)\right)^\alpha}}{\log\mean{\mu}{\left(1 + t\left(\frac{\dd\tilde{\nu}}{\dd\mu }-1\right)\right)^\alpha}}
    \end{equation}
    is increasing. 
    
    \textbf{Case $\alpha=2$:} Assume $\eta_2(\mu, K)\neq\eta_{\chi^2}(\mu, K)$ for otherwise there is nothing to prove. Note that this means the contraction of $K$ is non-trivial w.r.t. the $\chi^2$-Divergence since $\eta_{\chi^2}(\mu, K)<\eta_2(\mu, K)\leq1$. For $\alpha=2$, the function in~\cref{eq:function_to_show_monotonicity_of} (whose monotonicity we want to determine) simplifies to ${t\mapsto\frac{\log(1+t^2\chi^2(\tilde{\nu}K\|\mu K))}{\log(1+t^2\chi^2(\tilde{\nu}\|\mu))}}$, and since $t\mapsto t^2$ is itself increasing, we can effectively consider the monotonicity of $s\mapsto\frac{\log(1+s\chi^2(\tilde{\nu}K\|\mu K))}{\log(1+s\chi^2(\tilde{\nu}\|\mu))}$. Its derivative being positive amounts to
    \begin{align*}
        \frac{\frac{\chi^2(\tilde{\nu} K\|\mu K)}{1+s\chi^2(\tilde{\nu} K\|\mu K)}\log\big(1+s\chi^2(\tilde{\nu}\|\mu)\big) - \frac{\chi^2(\tilde{\nu}\|\mu)}{1+s\chi^2(\tilde{\nu}\|\mu)}\log\big(1+s\chi^2(\tilde{\nu} K\|\mu K)\big)}{\Big(\log\big(1+s\chi^2(\tilde{\nu}\|\mu)\big) \Big)^2} > 0\\
        \iff  \frac{\big(1+s\chi^2(\tilde{\nu} K\|\mu K)\big)\log\big(1+s\chi^2(\tilde{\nu} K\|\mu K)\big)}{\chi^2(\tilde{\nu} K\|\mu K)}< \frac{\big(1+s\chi^2(\tilde{\nu}\|\mu)\big)\log\big(1+s\chi^2(\tilde{\nu}\|\mu)\big)}{\chi^2(\tilde{\nu}\|\mu)}.
    \end{align*}
    Since the function $x\mapsto \frac{(1+sx)\log(1+sx)}{x}$ is increasing for $s\in(0, 1]$, the result holds because $\chi^2(\tilde{\nu} K\|\mu K)<\chi^2(\tilde{\nu}\|\mu)$ due to $\eta_{\chi^2}(\mu, K)<1$.

    \textbf{Case $\alpha\to\infty$:} Assume $\eta_\infty(\mu, K)\neq\eta_{L^\infty}(\mu, K)$ for otherwise there is nothing to prove. Note that this means the contraction of $K$ is non-trivial w.r.t. the $L^\infty$-Divergence since $\eta_{L^\infty}(\mu, K)<\eta_\infty(\mu, K)\leq1$ by applying~\cref{lemma:sdpi_relationship_concave_function} with ${g(x)=\log(1+x)}$. For $\alpha\to\infty$, the function in~\cref{eq:function_to_show_monotonicity_of} (whose monotonicity we want to determine) simplifies to $t\mapsto\frac{\log\left(1+t\left(\|\frac{\dd\tilde{\nu}K}{\dd\mu K}\|_{L^\infty(\mu K)}-1\right)\right)}{\log\left(1+t\left(\|\frac{\dd\tilde{\nu}}{\dd\mu}\|_{L^\infty(\mu)}-1\right)\right)}$ and we notice that the analysis thus becomes identical to the case $\alpha=2$, except that we now use the fact that ${\|\frac{\dd\tilde{\nu }K}{\dd\mu K}\|_{L^\infty(\mu K)}-1<\|\frac{\dd\tilde{\nu}}{\dd\mu}\|_{L^\infty(\mu)}-1}$.
    \end{IEEEproof}
\subsection{Proof of~\cref{prop:non_trivial_sdpi_characterisation}}\label{appendix:proof_non_trivial_sdpi_characterisation}\noindent
    We start with Part a). Since the boundary cases $\alpha\to0,1$ are already known~\cite{cohen1993relative}, let us consider $\alpha\in(0,1)$. 
    \begin{itemize}
        \item \textbf{If direction $\Longleftarrow$:}\\Suppose that there exist $x, x^\prime$ such that $\supp\big(K(\cdot|x)\big)$ and $\supp\big(K(\cdot|x^\prime)\big)$ are disjoint. Then~\cite[Theorem 4.1]{cohen1993relative} gives $\eta_{\chi^2}(K)=1$, and since $\eta_\alpha(K)\geq\eta_{\chi^2}(K)$ this means $\eta_\alpha(K)=1$.
        \item \textbf{Only if direction $\Longrightarrow$:}\\We prove the contrapositive. Assume that for all $x, x^\prime\in\X$, $\supp\big(K(\cdot|x\big)$ and $\supp\big(K(\cdot|x^\prime)\big)$ intersect. But then using that $\|\nu-\mu\|_{\rm TV}<1\iff \supp(\nu)\cap\supp(\mu)\neq\emptyset$ together with the characterisation ${\eta_{\rm TV}(K) = \sup_{x, x^\prime\in\X}\|K(\cdot|x)-K(\cdot|x^\prime)\|_{\rm TV}}$ gives $\eta_{\rm TV}(K)<1$. It remains to invoke Part a) of~\cref{prop:bound_on_distribution_independent_sdpi} and we get $\eta_\alpha(K)\leq\eta_{\rm TV}(K)<1$.
    \end{itemize}
    Let us now proceed with Part b) and analyse the case $\alpha>1$.
    \begin{itemize}
        \item \textbf{If direction $\Longleftarrow$:}\\Suppose that there exist two rows in $K$ with different support. In other words, there exists $x, x^\prime\in\X$ and $y\in\Y$ such that $K(y|x^\prime)=0$ but $K(y|x)>0$. Letting $\nu = \delta_x$ and $\mu_\varepsilon = (1-\varepsilon)\delta_{x^\prime}+\varepsilon\delta_x$ where $\varepsilon\in(0,1)$, we find that $\renyiDiv{\nu}{\mu_\varepsilon}=-\log\varepsilon$ and
        \begin{align*}
            \renyiDiv{\nu K}{\mu_\varepsilon K} &= \frac1{\alpha-1}\log\left(\sum_{y^\prime\in\Y}K (y^\prime|x)^\alpha\big((1-\varepsilon)K(y^\prime|x^\prime) +\varepsilon K(y^\prime|x)\big)^{1-\alpha}\right)\\
            &\geq \frac1{\alpha-1}\log\left(K (y|x)^\alpha\big((1-\varepsilon)\cdot 0 +\varepsilon K(y|x)\big)^{1-\alpha}\right)\\
            &= -\log\varepsilon -\frac1{\alpha-1}\log(K(y|x)).
        \end{align*}
        Consequently for all $\varepsilon\in(0,1)$,
        \begin{equation*}
            \eta_\alpha(K) \geq \frac{\renyiDiv{\nu K}{\mu_\varepsilon K}}{\renyiDiv{\nu}{\mu_\varepsilon}}= 1 -\frac{\log K(y|x)}{(\alpha-1)\log\frac1\varepsilon} \overset{\varepsilon\to 0}{\longrightarrow} 1,
        \end{equation*}
        and since $\eta_\alpha(K)\leq 1$ always, $\eta_\alpha(K)=1$.
        \item \textbf{Only if direction $\Longrightarrow$:}\\We prove the contrapositive. Assume that for all $x, x^\prime\in\X$, $\supp(K(\cdot|x)=\supp(K(\cdot|x^\prime))\equiv \calS$. Let us first cover the case where $\nu\not\ll\mu$, which could be obtained in a limiting case. In that case, we have $\renyiDiv{\nu}{\mu}=\infty$ while 
        \begin{align*}
            \renyiDiv{\nu K}{\mu K}&\leq D_\infty(\nu K\|\mu K) \\
            &= \log\max_{y\in\supp(\mu K)}\frac{\nu K(y)}{\mu K (y)}\\
            &\leq \log\max_{y\in\calS}\frac{\max_{x\in\X}K(y|x)}{\min_{x^\prime\in\X}K(y|x^\prime)} \\
            &<\infty.
        \end{align*}
        Such a limiting case would yield a ratio $\frac{\renyiDiv{\nu K}{\mu K}}{\renyiDiv{\nu}{\mu}} = 0$, which is obviously strictly smaller than 1. A second limiting case is when $\renyiDiv{\nu}{\mu}\to0$ (or equivalently $\nu\to \mu$), which is known to yield the local contraction $\eta_{\chi^2}(K)$~\cite[Theorem 2]{jin2024properties}. Since we assume all rows of $K$ have identical support, they are in particular not disjoint, so that $\eta_{\chi^2}(K)<1$. With the limiting cases out of the way, assume we have $0<\renyiDiv{\nu}{\mu}<\infty$. Using the identity in~\cref{eq:dual_kernel_formula}, one obtains
        \begin{align*}
            e^{(\alpha-1)\renyiDiv{\nu K}{\mu K}} &= \sum_{y\in\Y}\left(\dualK\frac{\dd\nu}{\dd\mu}(y)\right)^\alpha\mu K(y)\\
            &= \sum_{y\in\Y}\left(\sum_{x\in\X}\dualK(x|y)\frac{\nu(x)}{\mu(x)}\right)^\alpha\mu K(y)\\
            &<\sum_{y\in\Y}\sum_{x\in\X}\dualK(x|y)\left(\frac{\nu(x)}{\mu(x)}\right)^\alpha\mu K(y)\\
            &= \sum_{x\in\X}\left(\frac{\nu(x)}{\mu(x)}\right)^\alpha\sum_{y\in\Y}\dualK(x|y)\mu K(y)\\
            &=\sum_{x\in\X}\left(\frac{\nu(x)}{\mu(x)}\right)^\alpha\sum_{y\in\Y}K(y|x)\mu (x)\\
            &= e^{(\alpha-1)\renyiDiv{\nu}{\mu}},
        \end{align*}
        where the strict inequality follows from Jensen's inequality and the fact that $\frac{\dd\nu}{\dd\mu}$ is not constant. Combining the three cases ($\renyiDiv{\nu}{\mu}\to0,\infty$ and $0<\renyiDiv{\nu}{\mu}<\infty$), we conclude $\eta_\alpha(K)<1$.
    \end{itemize}
\subsection{Proof of~\cref{prop:distribution_independent_sdpi_equal_chi_square_sdpi}}\label{appendix:proof_distribution_independent_sdpi_equal_chi_square_sdpi}\noindent
    For the boundary cases ${\alpha\to0,1}$, we recover the distribution-independent SDPI constant of KL Divergence, an instance of $\varphi$-Divergence with operator convex $\varphi$, for which it is known that $\eta_{\varphi}(K)=\eta_{\chi^2}(K)$~\cite[Corollary III.1]{raginsky2016strong}. When $\alpha\in(0,1)$, the lower bound $\eta_\alpha(K)\geq \eta_{\chi^2}(K)$ has already been established~\cite[Corollary 2]{jin2024properties}. To prove the reverse inequality, we leverage the variational representation ${\renyiDiv{\nu}{\mu} = \inf_{\gamma\in\calP(\X)} \left\{\frac{\alpha}{1-\alpha}D_{\rm KL}(\gamma\|\nu) + D_{\rm KL}(\gamma\|\mu)\right\}}$ mentioned in~\cite[Theorem 30]{van2014renyi} (see also~\cite[Lemma II.1]{espositoGI2025sibson}). In particular, we find
    \begin{align*}
        \renyiDiv{\nu K}{\mu K} &=\inf_{\xi\in\calP(\Y)} \left\{\frac{\alpha}{1-\alpha}D_{\rm KL}(\xi\|\nu K) + D_{\rm KL}(\xi\|\mu K)\right\}\\
        &\leq \inf_{\gamma\in\calP(\X)} \left\{\frac{\alpha}{1-\alpha}D_{\rm KL}(\gamma K\|\nu K) + D_{\rm KL}(\gamma K\|\mu K)\right\}\\
        &\leq \inf_{\gamma\in\calP(\X)} \bigg\{\frac{\alpha}{1-\alpha}\eta_{\rm KL}(K)D_{\rm KL}(\gamma\|\nu) + \eta_{\rm KL}(K)D_{\rm KL}(\gamma\|\mu)\bigg\}\\
        &=\eta_{\rm KL}(K)\renyiDiv{\nu}{\mu}.
    \end{align*}
    We thus get $\eta_\alpha(K)\leq\eta_{\rm KL}(K)=\eta_{\chi^2}(K)$ and the conclusion follows.
\subsection{Proof of~\cref{prop:infty_sdpi_achieved_on_conditionals}}\label{appendix:proof_infty_sdpi_achieved_on_conditionals}\noindent
    Let us rewrite the optimisation problem equivalently as
    \begin{equation*}
        \eta_\infty(\mu, K) = \sup_{\substack{\nu\in\calP(\X)\\0<D_\infty(\nu\|\mu)<\infty}}\frac{D_\infty(\nu K\|\mu K)}{D_\infty(\nu\|\mu)} = \sup_{t>0}\frac{\sup_{\nu\in\calP_t} D_\infty(\nu K\|\mu K)}{t},
    \end{equation*}
    where $\calP_t=\{\nu\in\calP(\X): D_\infty(\nu\|\mu)\leq t\}$. Focusing on the inner supremisation, we can write
    \begin{align*}
        \sup_{\nu\in\calP_t} D_\infty(\nu K\|\mu K)
        &= \sup_{\nu\in\calP_t} \log\max_{y\in\supp(\mu K) }\frac{\nu K(y)}{\mu K(y)}\\
        &= \max_{y\in\supp(\mu K) }\log\sup_{\nu\in\calP_t}\frac{\nu K(y)}{\mu K(y)}\\
        &= \max_{y\in\supp(\mu K) }\log\frac{\sup_{\nu\in\calP_t}\sum_{x\in\X} \nu(x)K(y|x)}{\mu K(y)},
    \end{align*}
    where the swapping of the supremum with the maximum is allowed since the spaces have a finite number of elements and the logarithm is an increasing function. For any fixed $y$, notice how the inner supremisation can be seen as the following Fractional Knapsack Problem:
    \begin{align*}
        \text{Maximise }&\quad\sum_{x\in\X}  \nu(x)K(y|x)\\
        \text{Subject to }&\quad \nu(x)\leq e^t\mu(x)\quad \forall x\in\X,\\&\quad\sum_{x\in\X}\nu(x)=1.
    \end{align*}
    Writing $n=|\X|$, solving such a problem can be done in a greedy manner:
    \begin{itemize}
        \item Index the elements of $\X$ as $x_1, \dots,x_n$ so that $K(y|x_1)\geq K(y|x_2)\geq\dots \geq K(y|x_n)$.
        \item To maximise the linear objective, the optimal $\nu$ should put as much mass as possible for each index according to the established order while ensuring $\nu(x_i)\leq e^t\mu(x_i)$, and at the same time make sure that the constraint $\sum_{x\in\X}\nu(x)=1$ is respected. In other words, denoting by $k$ the largest index such that $\sum_{i=1}^k e^t\mu(x_i) \leq 1$, the optimal solution $\nu^\star$ is such that $\nu^\star(x_i)=e^t\mu(x_i)$ for $i=1,\dots,k$ and $\nu^\star(x_{k+1}) = 1-\sum_{i=1}^k e^t\mu(x_i)$ and $\nu^\star(x_j)=0$ for $j\geq k+1$.
    \end{itemize}
    For arbitrary fixed $y$ and $t$, the argument above reveals that $\nu^\star$ is in fact close to being a conditional distribution of $\mu$, which happens when the ``remainder'' term $\nu^\star(x_{k+1})$ is either 0 or equal to $e^t\mu(x_{k+1})$. Indeed, using the notation $\X_k\triangleq\{x_1, \dots, x_k\}$, if $\nu^\star(x_k)=0$ then we can write $\nu^\star(x)=\frac{\mu(x)\mathbbm{1}_{\X_k}(x)}{\mu(\X_k)}$, while if $\nu^\star(x_{k+1})=e^t\mu(x_k)$ we can write $\nu^\star(x)=\frac{\mu(x)\mathbbm{1}_{\X_{k+1}}(x)}{\mu(\X_{k+1})}$, which are both conditional distributions w.r.t. $\mu$. In the remainder of the proof, we show that one of those two cases is bound to happen.
    
    A solution where the ``remainder'' term $\nu^\star(x_{k+1})$ is at index $k+1$, must be such that $\sum_{i=1}^k e^t\mu(x_i) \leq 1$ and $1-\sum_{i=1}^k e^t\mu(x_i)=\nu^\star(x_{k+1})\leq e^t\mu(x_{k+1})$, from which one deduces $t\in\left[\log\frac1{\mu(\X_{k+1})}, \log\frac1{\mu(\X_k)}\right]$. Note that $\nu^\star$ is a conditional distribution w.r.t. $\mu$ when $t$ is at one of the boundaries. Moreover when the solution has the remainder term at index $k+1$, the function being maximised takes the value
    \begin{align*}
        \sum_{x\in\X}  \nu^\star(x)K(y|x) &= e^t\sum_{x\in\X_k}\mu(x)K(y|x) + \left(1-e^t\sum_{x\in\X_k}\mu(x)\right)K(y|x_{k+1})\\
        &=e^t\underbrace{\left(\sum_{x\in\X_k}\mu(x)\big(K(y|x)-K(y|x_{k+1})\big)\right)}_{\triangleq a_y\geq 0}+\underbrace{K(y|x_{k+1})}_{\triangleq b_y\geq 0},
    \end{align*}
    and consequently the overall problem takes the form
    \begin{equation*}
        \sup_{t\in\left[\log\frac1{\mu(\X_{k+1})}, \log\frac1{\mu(\X_k)}\right]}\frac{\max_{y\in\supp(\mu K)}\log(a_y e^t+b_y)}{t} = \max_{y\in\supp(\mu K)}\sup_{t\in\left[\log\frac1{\mu(\X_{k+1})}, \log\frac1{\mu(\X_k)}\right]}\frac{\log(a_y e^t+b_y)}{t}.
    \end{equation*}
    For $a_y, b_y\geq 0$, the function $t\mapsto \log(a_y e^t+b_y)$ is convex for $t>0$ and therefore $t\mapsto\frac{\log(a_y e^t+b_y)}{t}$ is quasiconvex~\cite[Example 3.38]{boyd2004convex}. Hence, its maximum over any interval must occur at an endpoint, and it follows that $\nu^\star$ must be a conditional distribution w.r.t. $\mu$. Lastly, since restricting $\mu$ to $A=\X$ gives back $\mu_{|A}=\mu$, it should not be considered in the optimisation defining $\eta_\infty(\mu, K)$. Consequently, we obtain the final expression
    \begin{equation*}
        \eta_\infty(\mu, K) = \sup_{A\subsetneq X}\frac{D_\infty\left(\mu_{|A} K\|\mu K\right)}{D_\infty\left(\mu_{|A}\|\mu\right)}.
    \end{equation*}
\subsection{Proof of~\cref{prop:bounds_on_renyi_infty_sdpi_by_tv}}\label{appendix:proof_bounds_on_renyi_infty_sdpi_by_tv}\noindent
    The result follows after upper and lower bounding the $\infty$-R\'enyi Divergence by the Total Variation Distance. First, notice that the following identity holds:
    \begin{equation*}
        \left\|\frac{\dd\nu}{\dd\mu}\right\|_{L^\infty(\mu)}= \sup_{\substack{A\in\Sigma_\X:\\\mu(A)>0}} \frac{\nu(A)}{\mu(A)}.
    \end{equation*}
    Let us also recall that ${\|\nu-\mu\|_{\rm TV} \triangleq \sup_{A\in\Sigma_\X}|\nu(A)-\mu(A)|}$. In particular for $\nu\ll\mu$, one can write ${\|\nu-\mu\|_{\rm TV} = \sup_{A\in\Sigma_\X:\mu(A)>0}\{\nu(A)-\mu(A)\}}$ and so in this case
    \begin{align*}
        D_\infty(\nu\|\mu) &= \log\left(\sup_{\substack{A\in\Sigma_\X:\\\mu(A)>0}} \frac{\nu(A)}{\mu(A)}\right)\\
        &\leq \sup_{\substack{A\in\Sigma_\X:\\\mu(A)>0}} \frac{\nu(A)}{\mu(A)}-1\\
        &\leq \frac1{\min_{x\in\supp(\mu)}\mu(x)}\|\nu-\mu\|_{\rm TV},
    \end{align*}
    where we used that $\log(x)\leq x-1$ for $x\geq 0$. Moreover, since $\log(x)\geq1-1/x$ for $x\geq 0$,
    \begin{align*}
        D_\infty(\nu\|\mu) &= \log\left(\sup_{\substack{A\in\Sigma_\X:\\\mu(A)>0}} \frac{\nu(A)}{\mu(A)}\right)\\
        &\geq \sup_{\substack{A\in\Sigma_\X:\\\mu(A)>0}} \left\{1-\frac{\mu(A)}{\nu(A)}\right\}\\
        &\geq \|\nu-\mu\|_{\rm TV},
    \end{align*}
    where in the last inequality we used $1/\nu(A)\geq 1$. The bounds can now be stated, but it is important to observe the set inclusion
    \begin{align}
        \{\nu\in\calP(\X): 0<D_\infty(\nu\|\mu)<\infty\} &= \{\nu\in\calP(\X): \nu\ll\mu \text{ and } \nu\neq\mu\}\nonumber \\&\subseteq \{\nu\in\calP(\X): \nu\neq\mu\}\label{eq:set_inclusion} \\&= \{\nu\in\calP(\X): 0<\|\nu-\mu\|_{\rm TV}<\infty\},\nonumber
    \end{align}
    due to the fact that Total Variation Distance is always bounded. Using this and the bounds just derived leads to
    \begin{align*}
        \eta_\infty(\mu, K) &= \sup_{\substack{\nu\in\calP(\X):\\\nu\neq\mu\\\nu\ll\mu}} \frac{D_\infty(\nu K\|\mu K)}{D_\infty(\nu\|\mu)} \\
        &\leq \frac1{\min_{y\in\supp(\mu K)}\mu K(y)}\cdot\sup_{\substack{\nu\in\calP(\X):\\\nu\neq\mu\\\nu\ll\mu}} \frac{\|\nu K-\mu K\|_{\rm TV}}{\|\nu-\mu\|_{\rm TV}} \\
        &\leq \frac1{\min_{y\in\supp(\mu K)}\mu K(y)}\cdot\sup_{\substack{\nu\in\calP(\X):\\\nu\neq\mu}} \frac{\|\nu K-\mu K\|_{\rm TV}}{\|\nu-\mu\|_{\rm TV}} \\
        &= \frac{1}{\min_{y\in\supp(\mu K)}\mu K(y)} \eta_{\rm TV}(\mu, K).
    \end{align*}
    The lower bound is deduced similarly, but the assumption that $\mu$ has full-support is needed to ensure that the set inclusion in~\cref{eq:set_inclusion} becomes an equality. In that case, we obtain
    \begin{align*}
        \eta_\infty(\mu, K) &= \sup_{\substack{\nu\in\calP(\X):\\\nu\neq\mu}} \frac{D_\infty(\nu K\|\mu K)}{D_\infty(\nu\|\mu)} \\
        &\geq \min_{x\in\supp(\mu)}\mu(x)\cdot\sup_{\substack{\nu\in\calP(\X):\\\nu\neq\mu}} \frac{\|\nu K-\mu K\|_{\rm TV}}{\|\nu-\mu\|_{\rm TV}} \\
        &= \min_{x\in\X}\mu(x)\cdot \eta_{\rm TV}(\mu, K).
    \end{align*}
\subsection{Proof of~\cref{prop:infty_renyi_sdpi_closed_form}}\label{appendix:proof_infty_renyi_sdpi_closed_form}\noindent
    By~\cref{prop:eta_parameter_renyi_binary} and~\cref{prop:infty_sdpi_achieved_on_conditionals}, it suffices to optimise over pairs of the form $\nu = \delta_x$ and $\mu_\varepsilon = (1-\varepsilon)\delta_x + \varepsilon\delta_{x^\prime}$ for some $x, x^\prime\in\X$, where $\varepsilon\in(0,1)$. The corresponding ratio of $\infty$-R\'enyi Divergences is given by
    \begin{align*}
        \frac{D_\infty(\delta_x K\|\mu_\varepsilon K)}{D_\infty(\delta_x\|\mu_\varepsilon)} &= \frac{\log \left\|\frac{\dd K(\cdot|x)}{\dd\mu_\varepsilon K}\right\|_{L^\infty(\mu_\varepsilon K)}}{\log\frac1{1-\varepsilon}}\\
        &= \frac{\log \max_{y\in\supp(\mu_\varepsilon K)}\frac{K(y|x)}{(1-\varepsilon)K(y|x) +\varepsilon K(y|x^\prime)}}{\log\frac1{1-\varepsilon}}\\
        &= \frac{\log \max_{y\in\supp(K(\cdot|x))}\frac{K(y|x)}{(1-\varepsilon)K(y|x) +\varepsilon K(y|x^\prime)}}{\log\frac1{1-\varepsilon}}\\
        &= \frac{\log \frac1{1-\varepsilon + \varepsilon \min_{y\in\supp(K(\cdot|x))}\frac{K(y|x^\prime)}{K(y|x)}}}{\log\frac1{1-\varepsilon}}\\
        &= \frac{\log \left(1-\varepsilon \left(1 -\min_{y\in\supp(K(\cdot|x))}\frac{K(y|x^\prime)}{K(y|x)}\right)\right)}{\log(1-\varepsilon)}.
    \end{align*}
    Now that the ratio has been determined, it remains to find its maximum value over $\varepsilon\in(0,1)$. To do so, we show that the map $\varepsilon \mapsto \frac{\log(1-c\varepsilon)}{\log(1-\varepsilon)}$ for $c\in[0,1]$ is decreasing, and we subsequently obtain the maximum by taking $\varepsilon\to 0$. Writing $u(\varepsilon) = -\log(1-\varepsilon)$, the map $u\mapsto \frac{g(u)}{u}$ over $u\geq 0$ can instead be analysed, with $g(x) = -\log(1 -c+ce^{-x})$. The derivative $g^\prime(x) = \frac1{1+\frac{1-c}{c}e^{x}}$ is decreasing so that $g$ is concave, and since $g(0)=0$ the same argument as in~\cref{appendix:sdpi_relationship_concave_function} gives that $\frac{g(u)}{u}$ is decreasing. Thus, the map $\varepsilon \mapsto \frac{\log(1-c\varepsilon)}{\log(1-\varepsilon)}$ is decreasing on $(0,1)$, and its maximum is attained when $\varepsilon\to 0$, which is equal to
    \begin{equation*}
        \lim_{\varepsilon\to 0} \frac{\log\left(1-c\varepsilon\right)}{\log(1-\varepsilon)} = \lim_{\varepsilon\to 0}\frac{\frac{-c}{1-c\varepsilon}}{\frac{-1}{1-\varepsilon}} = c.
    \end{equation*}
    In our case, this means $\sup_{\varepsilon\in(0,1)} \frac{D_\infty(\delta_x K\|\mu_\varepsilon K)}{D_\infty(\delta_x\|\mu_\varepsilon)} = 1 -\min_{y\in\supp(K(\cdot|x))} \frac{K(y|x^\prime)}{K(y|x)}$. Since $x, x^\prime\in\X$ are arbitrary, the final expression for the SDPI constant becomes
    \begin{equation*}
        \eta_\infty(\mu, K) = \sup_{x, x^\prime\in\X}\left\{1 -\min_{y\in\supp(K(\cdot|x))} \frac{K(y|x^\prime)}{K(y|x)}\right\}.
    \end{equation*}
\subsection{Details for the Comparison in~\cref{section:mixing_time}}\label{appendix:mixing_time_experiment}\noindent
    The Markov kernel under consideration is the tensor product of 3 Binary Symmetric Channels. That is, ${K=\mathrm{BSC}(\varepsilon)\otimes \mathrm{BSC}(\varepsilon)\otimes \mathrm{BSC}(\varepsilon)}$ with $\varepsilon=10^{-2}$, where ``$\otimes$'' denotes the tensor product. In this setting, one can view $\X$ as being the space $\X=\{0,1\}^3$ with 8 elements, and $K$ is more explicitly written as the $8\times8$ matrix
    \begin{equation*}
    K = \begin{bmatrix} 
        (1-\varepsilon)^3 & (1-\varepsilon)^2\varepsilon & (1-\varepsilon)^2\varepsilon & (1-\varepsilon)\varepsilon^2 & (1-\varepsilon)^2\varepsilon & (1-\varepsilon)\varepsilon^2 & (1-\varepsilon)\varepsilon^2 & \varepsilon^3 \\
        (1-\varepsilon)^2\varepsilon & (1-\varepsilon)^3 & (1-\varepsilon)\varepsilon^2 & (1-\varepsilon)^2\varepsilon & (1-\varepsilon)\varepsilon^2 & (1-\varepsilon)^2\varepsilon & \varepsilon^3 & (1-\varepsilon)\varepsilon^2 \\
        (1-\varepsilon)^2\varepsilon & (1-\varepsilon)\varepsilon^2 & (1-\varepsilon)^3 & (1-\varepsilon)^2\varepsilon & (1-\varepsilon)\varepsilon^2 & \varepsilon^3 & (1-\varepsilon)^2\varepsilon & (1-\varepsilon)\varepsilon^2 \\
        (1-\varepsilon)\varepsilon^2 & (1-\varepsilon)^2\varepsilon & (1-\varepsilon)^2\varepsilon & (1-\varepsilon)^3 & \varepsilon^3 & (1-\varepsilon)\varepsilon^2 & (1-\varepsilon)\varepsilon^2 & (1-\varepsilon)^2\varepsilon \\
        (1-\varepsilon)^2\varepsilon & (1-\varepsilon)\varepsilon^2 & (1-\varepsilon)\varepsilon^2 & \varepsilon^3 & (1-\varepsilon)^3 & (1-\varepsilon)^2\varepsilon & (1-\varepsilon)^2\varepsilon & (1-\varepsilon)\varepsilon^2 \\
        (1-\varepsilon)\varepsilon^2 & (1-\varepsilon)^2\varepsilon & \varepsilon^3 & (1-\varepsilon)\varepsilon^2 & (1-\varepsilon)^2\varepsilon & (1-\varepsilon)^3 & (1-\varepsilon)\varepsilon^2 & (1-\varepsilon)^2\varepsilon \\
        (1-\varepsilon)\varepsilon^2 & \varepsilon^3 & (1-\varepsilon)^2\varepsilon & (1-\varepsilon)\varepsilon^2 & (1-\varepsilon)^2\varepsilon & (1-\varepsilon)\varepsilon^2 & (1-\varepsilon)^3 & (1-\varepsilon)^2\varepsilon \\
        \varepsilon^3 & (1-\varepsilon)\varepsilon^2 & (1-\varepsilon)\varepsilon^2 & (1-\varepsilon)^2\varepsilon & (1-\varepsilon)\varepsilon^2 & (1-\varepsilon)^2\varepsilon & (1-\varepsilon)^2\varepsilon & (1-\varepsilon)^3 
        \end{bmatrix}.
    \end{equation*}
    The stationary distribution $\pi$ corresponding to this Markov chain is the uniform distribution over $\X$, that is $\pi=\begin{bmatrix}
        \frac18 &\cdots&\frac18
    \end{bmatrix}$, and we choose the ``starting'' distribution to be $\nu=\delta_{000}=\begin{bmatrix}
        1 & 0&\cdots&0
    \end{bmatrix}$. In order to calculate the bounds in~\cref{eq:mcmt_linear_sdpi,eq:mcmt_non_linear_sdpi}, both $\eta_{\chi^2}(\pi, K)$ and $\eta_2(\pi, K)$ need to be evaluated. Since the size of the space is not too large (namely $|\X|=8$), both constant can be computed numerically.
\end{document}